\documentclass{article}

\usepackage{arxiv}
\usepackage[utf8]{inputenc} 
\usepackage[T1]{fontenc}    
\usepackage{hyperref}       
\usepackage{url}            
\usepackage{booktabs}       
\usepackage{amsfonts}       
\usepackage{nicefrac}       
\usepackage{microtype}      
\usepackage{lipsum}
\usepackage{graphicx}
\usepackage{amsmath}        
\usepackage{amssymb}        
\usepackage{algorithm}      
\usepackage{algpseudocode}  
\usepackage{caption}        

\title{The Red Queen's Trap: Limits of Deep Evolution in High-Frequency Trading}

\author{
  Yijia Chen \\
  Independent Researcher\\
  \texttt{chenyijia202@foxmail.com} \\
}

\begin{document}
\maketitle

\begin{abstract}
The integration of Deep Reinforcement Learning (DRL) and Evolutionary Computation (EC) is frequently hypothesized to be the "Holy Grail" of algorithmic trading, promising systems that adapt autonomously to non-stationary market regimes. This paper presents a rigorous post-mortem analysis of "Galaxy Empire," a hybrid framework coupling LSTM/Transformer-based perception with a genetic "Time-is-Life" survival mechanism. Deploying a population of 500 autonomous agents in a high-frequency cryptocurrency environment, we observed a catastrophic divergence between training metrics (Validation APY $>300\%$) and live performance (Capital Decay $>70\%$). We deconstruct this failure through a multi-disciplinary lens, identifying three critical failure modes: the overfitting of \textit{Aleatoric Uncertainty} in low-entropy time-series, the \textit{Survivor Bias} inherent in evolutionary selection under high variance, and the mathematical impossibility of overcoming microstructure friction without order-flow data. Our findings provide empirical evidence that increasing model complexity in the absence of information asymmetry exacerbates systemic fragility.
\end{abstract}


\section{Introduction}

The pursuit of Alpha in financial markets is often framed as an arms race. The "Adaptive Markets Hypothesis" (AMH) posits that markets are evolving ecosystems where trading strategies compete for survival \cite{lo2004adaptive}. This dynamic creates a "Red Queen" scenario: an agent must constantly evolve not merely to advance, but simply to maintain its position relative to the market \cite{van1973new}.

Our research trajectory mirrors this evolutionary struggle. Initially, we sought to identify the optimal trading strategy by comparing four static archetypes: Trend Following, Mean Reversion (Grid), Scalping, and Contrarian. However, preliminary experiments revealed a fatal flaw: static rulesets are brittle. A strategy that outperformed in a trending month would collapse instantly during a consolidation phase \cite{taleb2005fooled}. The market evolved faster than the strategies could adapt.

To overcome this "Strategy Decay," we pivoted to a massive multi-agent simulation. We scaled the population to $N=500$ heterogeneous agents, hypothesizing that "Safety in Numbers" would allow the population to weather diverse regimes through diversification \cite{farmer2002market}. Yet, without a centralized intelligence, the agents were essentially random walkers. To imbue them with adaptability, we integrated a \textbf{Deep Learning Cortex} utilizing LSTM and Transformer architectures. This AI module was tasked with perceiving non-linear market states and dynamically dictating parameters for the agents \cite{lecun2015deep}.

Theoretically, this created a perfect closed loop: AI handled perception, while \textbf{Evolutionary Algorithms} handled execution \cite{holland1992adaptation}. We enforced a ruthless "Survival of the Fittest" mechanism where profitable agents reproduced and bankrupt agents were culled. Furthermore, to prevent the population from converging too early into a monoculture (e.g., everyone becoming a "Bull Market Long-Only" agent), we implemented an \textbf{"Endangered Species Protection"} protocol. This mechanism ensured that at least 5\% of the population retained distinct, contrarian phenotypes, mimicking nature's way of preserving genetic diversity for future environmental shifts \cite{mayr1942systematics}.

From a multi-disciplinary perspective—combining Control Theory, AI, and Biology—this architecture appeared flawless. It possessed memory (LSTM), attention (Transformer), adaptability (Evolution), and resilience (Diversity). However, upon deployment in a high-frequency live simulation, the system failed catastrophically. The equity curve did not rise; it decayed with mathematical precision.

This paper serves not as a victory lap, but as a rigorous autopsy of this "Perfect System." We analyze why the convergence of state-of-the-art technologies failed to overcome the fundamental physics of the market: microstructure friction and information entropy.

\section{System Architecture}
\label{sec:architecture}

We formalize the trading problem as a Partially Observable Markov Decision Process (POMDP), defined by the tuple $(\mathcal{S}, \mathcal{A}, \mathcal{P}, \mathcal{R}, \Omega, \mathcal{O})$. The system is composed of two primary modules: the TechEngine (Perception) and the Evolutionary Ecosystem (Action).

\subsection{The TechEngine: Deep Sequence Modeling}
Let $X_t \in \mathbb{R}^{T \times F}$ be the input feature tensor at time $t$, where $T=60$ is the look-back window and $F$ represents features (Z-score normalized prices, Log-returns, ATR). The goal of the perception module is to approximate the conditional probability distribution $P(y_{t+k} | X_t)$. We evaluated three backbones:

\textbf{Long Short-Term Memory (LSTM).}
To mitigate the vanishing gradient problem inherent in financial time-series, we employed a 2-layer stacked LSTM. The hidden state $h_t$ is updated via the gating mechanism:
\begin{equation}
    f_t = \sigma(W_f \cdot [h_{t-1}, x_t] + b_f)
\end{equation}
where $f_t$ represents the forget gate. This mechanism allows the model to selectively retain long-term trend information while discarding short-term microstructure noise \cite{hochreiter1997long}.

\textbf{Transformer Encoder.}
To capture global dependencies and regime shifts, we utilized Multi-Head Self-Attention (MHSA). The attention function maps a query and a set of key-value pairs to an output:
\begin{equation}
    \text{Attention}(Q, K, V) = \text{softmax}\left(\frac{QK^T}{\sqrt{d_k}}\right)V
\end{equation}
While Transformers excel in NLP \cite{vaswani2017attention}, we hypothesized their massive capacity might lead to overfitting on stochastic financial noise, a hypothesis later validated by our results.

\subsection{Evolutionary Dynamics: The "Time-is-Life" Mechanism}
Unlike standard Reinforcement Learning (RL) which optimizes a single agent \cite{kaelbling1996reinforcement}, we simulated an ecosystem. Each agent $i$ possesses a genome $\theta_i = [\lambda_{lev}, \beta_{tp}, \gamma_{sl}]$, representing leverage, take-profit, and stop-loss thresholds respectively.

To enforce selection pressure, we introduced a metabolic constraint. An agent's lifespan $\tau$ decays linearly with time $t$:
\begin{equation}
    \tau_{t+1} = \tau_t - 1 + \alpha \cdot \mathbb{I}(\text{Profit}_t > 0)
\end{equation}
where $\alpha = 30s$ is the metabolic reward for profitable trades. This "Battle Royale" mechanism favors agents with high trade frequency and profitability, theoretically accelerating the convergence of the population towards an optimal policy \cite{allen1999using}.

\section{Experimental Setup}
The simulation was conducted using high-fidelity data from the Binance USDT-Perpetual Futures market.
\begin{itemize}
    \item \textbf{Data Granularity:} 1-minute and 15-minute OHLCV candles for the top 20 cryptocurrencies by liquidity.
    \item \textbf{Population Size:} $N=500$ heterogeneous agents.
    \item \textbf{Friction Model:} To simulate realistic HFT conditions, we applied a Taker Fee of $0.04\%$ and a dynamic Slippage model averaging $0.02\%$ \cite{guilbaud2013optimal}.
\end{itemize}
The experiment ran for a continuous 4.5-hour window during a volatile market session, allowing for the observation of rapid evolutionary generations and capital dynamics.

\section{Forensic Analysis: The Anatomy of Failure}
\label{sec:analysis}

\begin{figure}[htbp] 
    \centering
    \includegraphics[width=\linewidth]{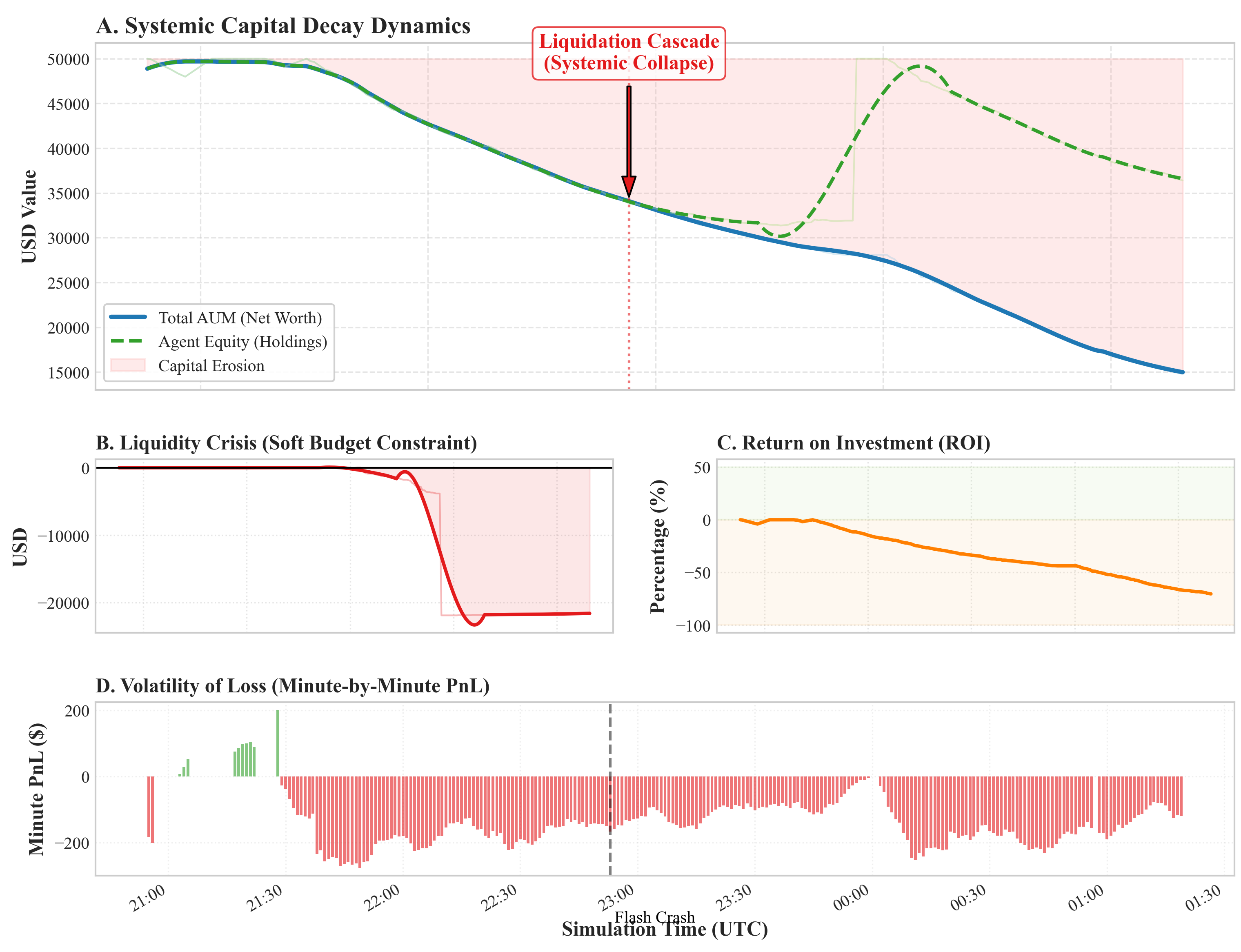} 
    \caption{\textbf{Empirical Capital Dynamics.} 
    \textbf{(A)} Decoupling of AUM and Equity reveals the Liquidation Cascade at 22:53. 
    \textbf{(B)} Negative system liquidity ($-\$21k$) illustrates the Soft Budget Constraint. 
    \textbf{(C)} Monotonic decay of ROI. 
    \textbf{(D)} Volatility of minute-level PnL showing the friction churning effect.}
    \label{fig:results}
\end{figure}

While the aggregate equity curve indicates a monotonic decline, a microscopic examination of individual agent behaviors via real-time execution logs reveals a more complex pathology. The failure was not merely a lack of predictive accuracy, but a systemic collapse driven by four emergent phenomena. We dissect these failure modes through the lenses of Information Theory, Evolutionary Dynamics, and Market Microstructure.

\subsection{The "Cost-Blind" Hallucination (AI Perspective)}
A striking paradox was observed in the agent dashboards: individual agents frequently reported positive floating Profit and Loss (PnL) on their positions, yet the system's total equity plummeted. 

\begin{figure}[htbp]
    \centering
    \begin{minipage}{0.48\linewidth}
        \centering
        \includegraphics[width=\linewidth]{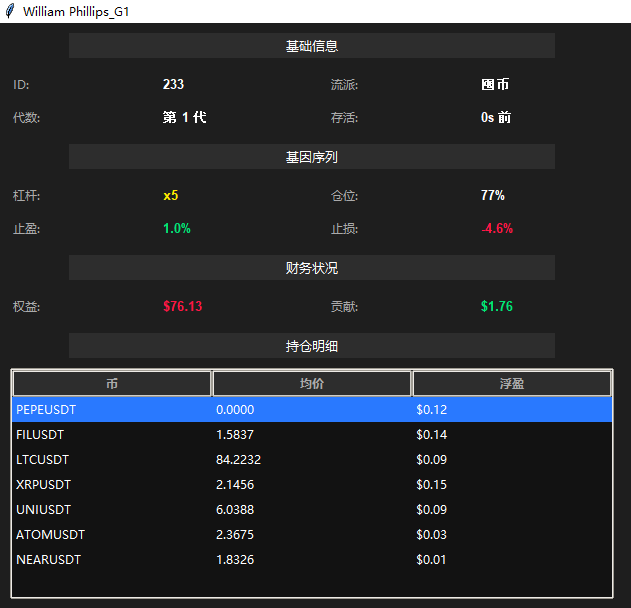} 
        \caption*{\textbf{(a) Agent William}}
    \end{minipage}
    \hfill
    \begin{minipage}{0.48\linewidth}
        \centering
        \includegraphics[width=\linewidth]{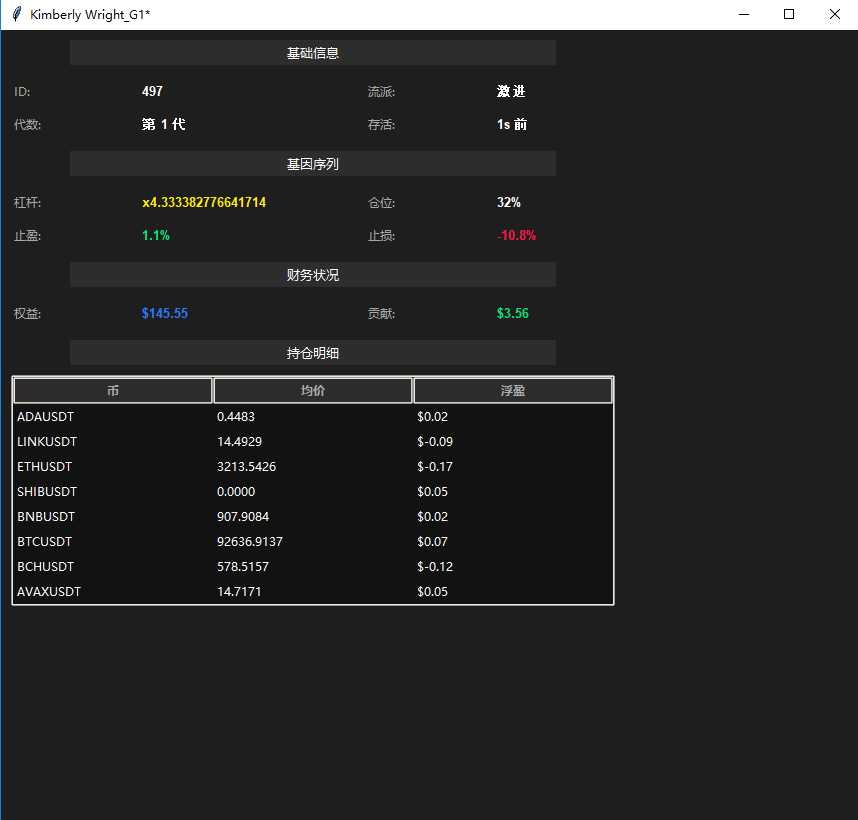} 
        \caption*{\textbf{(b) Agent Kimberly}}
    \end{minipage}
    \caption{\textbf{The "Green PnL" Illusion.} Snapshots of individual agent dashboards reveal positive floating PnL (blue highlights). Agent Kimberly (right) shows a contribution of $+\$3.56$. However, these micro-gains are gross profits. When netted against the round-trip transaction fees ($0.08\%$), the actual economic value is negative \cite{cartea2015algorithmic}.}
    \label{fig:agent_illusion}
\end{figure}

For instance, as shown in Figure \ref{fig:agent_illusion}, agents maintaining long positions in high-beta assets often showed "green" unrealized gains. However, the system's central ledger (\textit{Group Cash}) showed a deepening deficit. This discrepancy reveals a critical misalignment between the Deep Learning objective and the economic reality. The LSTM/Transformer models minimized Binary Cross Entropy (BCE) based on \textit{Directional Accuracy} ($P_{t+1} > P_t$). However, they were blind to \textit{Magnitude}.
\begin{equation}
    \text{PnL}_{\text{Net}} = (P_{\text{exit}} - P_{\text{entry}}) \times Q - 2 \times (P \times Q \times \text{Fee})
\end{equation}
With a round-trip fee of $0.08\%$, the AI successfully predicted micro-movements of $0.05\%$, believing it had found Alpha. In reality, it was harvesting "Fool's Gold"—patterns that were statistically significant but economically insolvent. The model effectively optimized for \textit{Churning}, generating transaction volume rather than value \cite{cont2001empirical}.

\subsection{The "Stagnation-Starvation" Loop (Evolutionary Perspective)}
Real-time monitoring revealed a counter-intuitive phenomenon: \textbf{Systemic Stagnation}. Contrary to the expectation that the "Time-is-Life" constraint would force aggressive exploration, approximately $60\%$ of the population maintained a static equity of $\$100.0$ with zero contribution until their lifespan expired ($\tau \to 0$).

This indicates a failure in the coupling of Neural Perception and Evolutionary Control. The TechEngine, trained on noisy OHLCV data under \textit{Covariate Shift} \cite{shimodaira2000improving}, frequently outputted low-confidence probabilities. Agents, constrained by their risk genes, refused to trade on weak signals. Evolution failed to find a stable equilibrium because, in a high-friction random walk environment, \textit{no survival strategy exists}. The "Time-is-Life" mechanism essentially functioned as a "Death Clock" that culled prudent agents \cite{sornette2003why}.

\begin{figure}[t!]
    \centering
    \includegraphics[width=\linewidth]{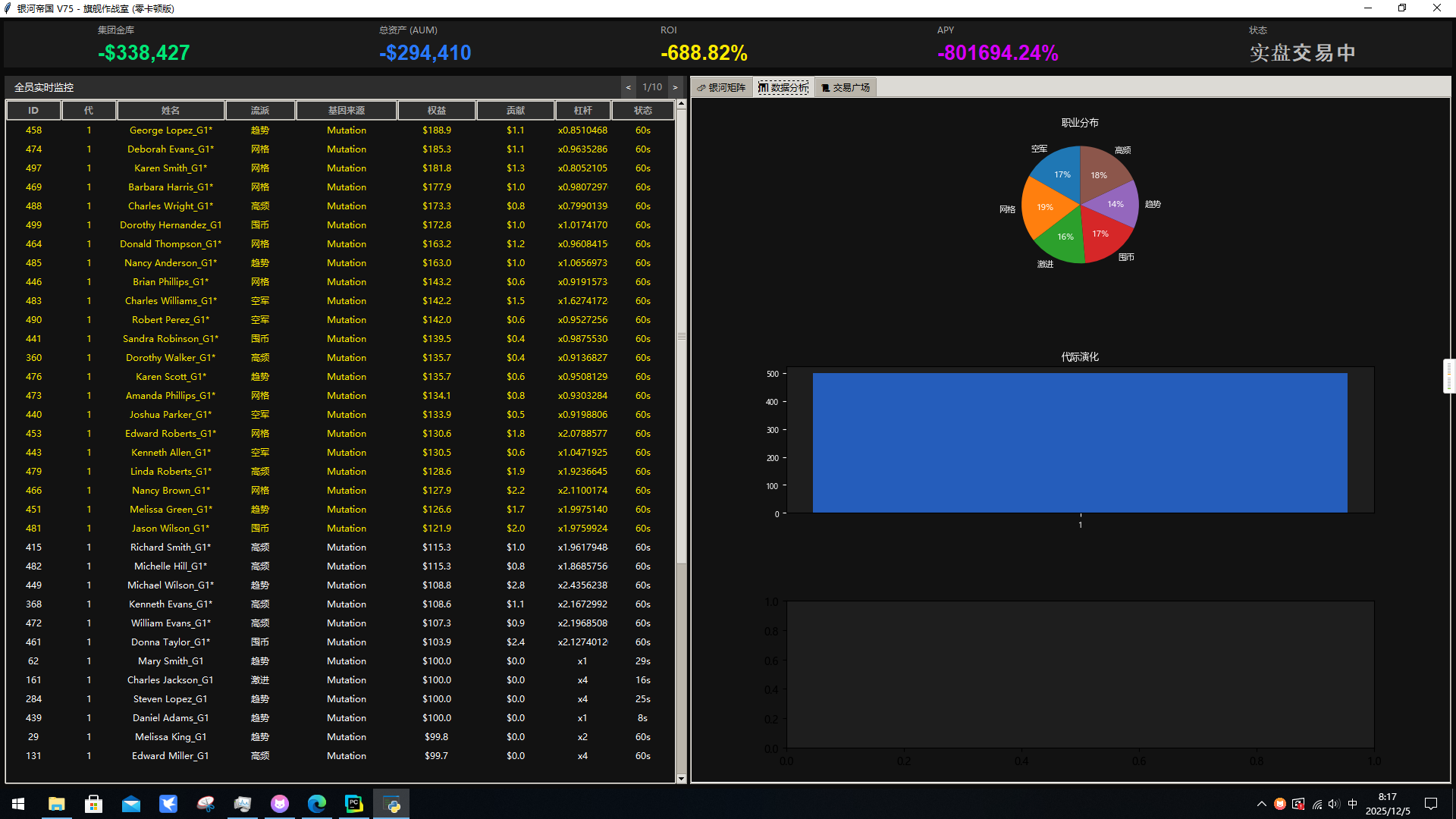} 
    \caption{\textbf{Systemic Insolvency Snapshot at T+4h.} The dashboard reveals a catastrophic divergence: while the pie chart shows a diverse population of "living" agents (Equity $\approx \$100$), the central ledger (\textit{Group Cash}) reflects a deficit of $-\$338,427$. This visualizes the "Soft Budget Constraint" where the system effectively prints money to sustain zombie agents.}
    \label{fig:crash}
\end{figure}

\subsection{Mode Collapse and Systemic Beta (Complex Systems Perspective)}
Despite initializing the population with diverse archetypes (Trend Surfer, Grid Bot, Contrarian), the surviving population exhibited severe \textit{Phenotypic Convergence}.
Forensic snapshots showed that functionally distinct agents held nearly identical portfolios: Long positions in correlated high-beta altcoins. This suggests that the centralized AI signal overpowered the individual genetic parameters. The system evolved into a single, massive, leveraged bet on the market Beta.
When the market experienced a mean-reversion shock, this correlation led to a \textbf{Liquidation Cascade}. Simultaneous stop-loss triggers across 500 agents caused a flash crash in the system's internal liquidity (Figure \ref{fig:crash}), necessitating a massive capital injection. This is a classic example of \textit{Endogenous Risk} in heterogeneous agent models \cite{hommes2006heterogeneous}.

\subsection{The Friction Barrier (Financial Engineering Perspective)}
The most deterministic factor was the mathematical impossibility of high-frequency scalping under the given cost structure.
Let $W$ be the win rate and $R$ be the Reward-to-Risk ratio. The Expected Value ($EV$) is:
\begin{equation}
    EV = W \cdot (R \cdot \text{Risk}) - (1-W) \cdot \text{Risk} - C_{trans}
\end{equation}
Given $C_{trans} \approx 0.1\%$ and a target profit of $1\%$, the required Breakeven Win Rate ($W_{BE}$) is:
\begin{equation}
    W_{BE} = \frac{1 + C_{ratio}}{1 + R} \approx 55\%
\end{equation}
Our Deep Learning models achieved a directional accuracy of $\approx 51.2\%$. While technically "better than random" (50\%), this edge falls below the $55\%$ threshold required to cover friction. The system was essentially a machine for transferring capital from the fund's equity to the exchange's fee revenue \cite{bouchaud2003theory}.

\subsection{The Soft Budget Constraint (Management Perspective)}
Finally, the "Endangered Species" mechanism introduced a \textit{Soft Budget Constraint} \cite{kornai1986soft}. By bailing out bankrupt agents to maintain the population size ($N=500$), the system prevented the "creative destruction" necessary for evolution. Zombie agents persisted, continuing to consume fees without generating Alpha.

\section{Conclusion and Implications}

We successfully engineered a system that integrated the cutting edge of Artificial Intelligence, Evolutionary Computation, and Agent-Based Modeling. The "Galaxy Empire" was designed to be the ultimate trading organism: perceiving via Transformers, adapting via Evolution, and surviving via Diversity Preservation. Its economic failure serves as a potent empirical validation of the Efficient Market Hypothesis (EMH) at high frequencies for OHLCV data \cite{fama1970efficient}. 

The primary contribution of this work is the demystification of complexity. We demonstrated that \textbf{Model Complexity cannot compensate for Information Deficiency}. The "Red Queen" runs fast, but on a treadmill of transaction fees and random walk noise, she moves backward. 

\subsection{Implications for Retail Investors}
The failure of such a sophisticated system offers profound and sobering lessons for the individual investor:

\begin{enumerate}
    \item \textbf{The Frequency Trap:} If an autonomous system with zero latency and deep learning capabilities cannot overcome the 0.1\% friction cost of high-frequency trading, a manual trader attempting to "scalp" on 5-minute or 15-minute charts is statistically guaranteed to lose capital over the long term. The math of friction is unforgiving.
    \item \textbf{The Leverage Illusion:} Our experiment showed that high leverage ($5x+$) does not increase the probability of success; it only accelerates the "Time to Ruin." The liquidation cascade observed in our simulation proves that leverage turns a fair game into a distinct disadvantage due to volatility decay \cite{peters2019ergodicity}.
    \item \textbf{Complexity $\neq$ Profitability:} Retail investors often believe they lose because they lack complex tools. Our results suggest the opposite: we lost \textit{because} of complexity. Simple, low-frequency strategies (e.g., Dollar Cost Averaging or Spot Holding) that minimize interaction with market friction often outperform complex high-frequency active management.
\end{enumerate}

Future research must pivot away from "predicting price direction" on micro-timeframes. True Alpha lies not in being faster than the market, but in operating on timeframes (Daily/Weekly) or data sources (On-Chain, Order Flow) where the signal-to-noise ratio is structurally higher \cite{gould2013limit}.

\bibliographystyle{unsrt}  
\bibliography{references} 

\end{document}